\documentclass[12pt]{report}
\usepackage[dvips]{graphicx}
\newcommand{\sptwo}{1.4}

\newcommand{\doublespace}{\edef\baselinestretch{\sptwo}\Large\normalsize}

\begin{document}
\doublespace

\begin{center}
{\bf Density Functional Theory of Bosons in a Trap}\\

\renewcommand\thefootnote{\fnsymbol{footnote}}
{Yeong E. Kim \footnote{ e-mail:yekim$@$physics.purdue.edu} and
Alexander L. Zubarev\footnote{ e-mail: zubareva$@$physics.purdue.edu}}\\
Purdue Nuclear and Many-Body Theory Group (PNMBTG)\\
Department of Physics, Purdue University\\
West Lafayette, Indiana  47907\\
\end{center}
\begin{quote}
 A time-dependent Kohn-Sham (KS) like theory is presented for $N$ bosons in three and
 lower-dimensional traps.  We derive coupled equations, which allow one to 
calculate the energies of elementary excitations. A rigorous proof is given
 to show that the KS like equation correctly describes properties of
 the  one-dimensional condensate
 of impenetrable bosons in a general time-dependent harmonic trap 
in the large $N$ limit. 
\end{quote}

\vspace{5mm}

\vspace{55 mm}
\noindent

\pagebreak

\vspace{8pt}

The newly created Bose-Einstein condensates (BEC) of weakly interacting 
alkali-metal atoms [1] stimulated a large number of theoretical investigations
 (see recent reviews [2]).
Most of these works are based on the assumption that  the  properties of BEC
are well described by the Gross-Pitaevskii (GP) mean-field theory [3].
The validity of the GP equation is nearly universally accepted.

The experimental realization of quasi-one dimensional (1D) and quasi-two dimensional (2D) trapped gases [4-6]
stimulated many theoretical interests.
The theoretical aspects of the BEC in quasi-1D and quasi-2D traps 
have been reported in many papers [7-17].
For the case of dimensions $d<3$, it is known that the quantum-mechanical two-body $t$-matrix 
vanishes
[18] at low energies. Therefore the replacement 
of the two-body interaction by the $t$-matrix, as done in deriving the GP mean-field theory,  is not correct in general for $d<3$ [12,19].

The density functional theory (DFT), originally developed for interacting 
systems of fermions [20], provides a rigorous alternative approach to the 
interacting inhomogeneous Bose gases.
The main goal of this letter is to develop a Kohn-Sham (KS) like time-dependent theory for bosons. 

We consider a system of $N$ interacting bosons in a trap potential $V_{ext}$. Assuming that our system is in a local thermal equilibrium at each position $\vec{r}$ with the local energy per particle $\epsilon(n)$ ($\epsilon$
is the ground state energy per particle of the homogeneous system and $n$ is the density), we can write a zero temperature classical hydrodynamics equation as 
[8]
$$
\frac{\partial n}{\partial t}+\nabla (n \vec{\mbox{v}})=0,
\eqno{(1)}
$$
$$
\frac{\partial \vec{\mbox{v}}}{\partial t}+\frac{1}{m}\nabla (V_{ext}+\frac{\partial(n \epsilon(n))}{\partial n}+\frac{1}{2} m \mbox{v}^2)=0,
\eqno{(2)}
$$
where $\vec{\mbox{v}}$ is the velocity field.

Adding the kinetic energy pressure term, we have
$$
\frac{\partial \vec{\mbox{v}}}{\partial t}+\frac{1}{m}\nabla (V_{ext}+\frac{\partial(n \epsilon(
n))}{\partial n}+\frac{1}{2} m \mbox{v}^2-\frac{\hbar^2}{2 m} \frac{1}{\sqrt{n}} \nabla^2 \sqrt{n})=0.\eqno{(3)}
$$
We define the density of the system as
$$
n(\vec{r},t)=\mid \Psi(\vec{r},t)\mid^2,
\eqno{(4)}
$$
and the velocity field $\vec{\mbox{v}}$ as
$$
\vec{\mbox{v}}(\vec{r},t)=\frac{\hbar}{2 i m n(\vec{r},t)}(\Psi^\ast \nabla \Psi-
\Psi \nabla \Psi^\ast).
\eqno{(5)}
 $$

From  Eqs.(1) and (5), we obtain  the following KS like time-dependent equation
$$
i\hbar \frac{\partial \Psi}{\partial t}=-\frac{ \hbar^2}{2 m} \nabla^2 \Psi
+V_{ext} \Psi+\frac{\partial( n \epsilon(n))}{\partial n}\Psi.
\eqno{(6)}
$$
If the trap potential, $V_{ext}$, is independent of time, one can write the 
ground-state wave function as $\Psi(\vec{r},t)=\Phi(\vec{r}) \exp(-i \mu t/\hbar)$,
where $\mu$ is the chemical potential, and $\Phi$ is normalized to the total number of particles, $\int d\vec{r} \mid \Phi \mid^2=N$. Then Eq.(6) becomes [21]
$$
(-\frac{ \hbar^2}{2 m} \nabla^2 +V_{ext}+\frac{\partial( n \epsilon(n))}
{\partial n})\Phi=\mu \Phi,
\eqno{(7)}
$$
where the solution of the equation (7) minimizes the KS energy functional in the local
 density approximation (LDA)
$$
E=N<\Phi\mid -\frac{ \hbar^2}{2 m} \nabla^2+V_{ext}+\epsilon(n)\mid \Phi>,
\eqno{(8)}
$$
and the chemical potential $\mu$ is given by $\mu=\partial E/\partial N$.
Eq.(7) has the form of the KS equation.

The ground-state energy per particle of the homogeneous system  $\epsilon(n)$
 for dilute $3D$ [22] and dilute $2D$ [23] Bose gases are
$$
\epsilon(n)=\frac{2 \pi \hbar^2}{m} a_{3D}n[1+\frac{128}{15 \sqrt{\pi}}
(n a_{3D}^3)^{1/2}+8 (\frac{4 \pi}{3}-\sqrt{3}) na_{3D}^3 \ln(n a_{3D}^3)+...],
\eqno{(9)}
$$
and
$$
\epsilon(n)=\frac{2 \pi \hbar^2 n}{m} \mid \ln(n a_{2D}^2)\mid^{-1} (1+O\mid \ln(n a_{2D}^2)\mid^{-1/5}),
\eqno{(10)}
$$
where $a_{3D}$ and $a_{2D}$ are $3D$ and $2D$ scattering lengths respectively.

For $1D$ Bose gas interacting via a repulsive $\delta$-function potential, $ \tilde{g} \delta(x)$, $\epsilon(n)$ is given by [24]
$$
\epsilon(n)=\frac{\hbar^2}{2 m} n^2 e(\gamma),
\eqno{(11)}
$$
where $\gamma=m\tilde{g}/(\hbar^2 n)$ and
$$
\everymath={\displaystyle}
\begin{array}{rcl}
e(\gamma)=\frac{\gamma^3}{\lambda^3}\int_{-1}^{+1}g(x) x^2 dx,\\ \\
g(y)=\frac{1}{2\pi}(1+2 \lambda \int_{-1}^{+1}\frac{g(x) dx}{\lambda^2+(x-y)^2}), \\ \\
\lambda=\gamma \int_{-1}^{+1}g(x) dx.
\end{array}
\eqno{(12)}
$$
For small values of $\gamma$, the following expression for $\epsilon(n)$,
$$
\epsilon(n)=\frac{\tilde{g}}{2} (n-\frac{4}{3 \pi}\sqrt{\frac{m\tilde{g} n}{\hbar^2}})+...
\eqno{(13)}
$$
is adequate up to approximately $\gamma=2$ [24].

For a large coupling strength $\tilde{g}$ [24]
$$
\epsilon(n)=\frac{\hbar^2 \pi^2 n^2}{6 m}(1+\frac{2 \hbar^2 n}{m\tilde{g}})^{-2}.
\eqno{(14)}
$$
 Eq.(14) is accurate to 1\% for $\gamma \geq 10$ [24].

In the limit of large $N$, by neglecting the kinetic energy term in the KS equation (7), we obtain an equation corresponding to the Thomas-Fermi (TF) approximation
$$
V_{ext}+\frac{\partial( n \epsilon(n))}
{\partial n}=\mu
\eqno{(15)}
$$
in the region where $n(\vec{r})$ is positive and $n(\vec{r})=0$ outside this region.

Now we turn our attention to  elementary excitations, corresponding to  small oscillations of  $\Psi(\vec{r},t)$ around the ground state.
Elementary excitations can be obtained by standard linear response analysis [25,26] of the equation (6), as resonances in the linear response.
 We add a weak sinusoidal perturbation to the time-dependent equation (6)
$$
i\hbar \frac{\partial \Psi}{\partial t}=(-\frac{ \hbar^2}{2 m} \nabla^2 
+V_{ext} +\frac{\partial( n \epsilon(n))}{\partial n}+f_+e^{-i \omega t}+
f_- e^{i \omega t})\Psi,
\eqno{(16)}
$$
 and assume that solution of Eq.(16) has the following form
$$
\Psi(\vec{r},t)=e^{-i \mu t/\hbar}[\Phi(\vec{r})+u(\vec{r}) e^{-i \omega t}+
v^\ast (\vec{r}) e^{i\omega t}],
\eqno{(17)}
$$
where $\Phi(\vec{r})$ is the ground-state solution of Eq.(7).

Linearization in  small amplitudes $u$ and $v$ yields inhomogeneous equations
$$
\everymath={\displaystyle}
\begin{array}{rcl}
(L-\hbar \omega)u+\frac{\partial^2(n\epsilon(n))}{\partial n^2} \Phi^2 v=-f_+\Phi,\\ \\
(L+\hbar \omega)v+\frac{\partial^2(n\epsilon(n))}{\partial n^2} \Phi^{\ast 2}u=-f_-\Phi,
\end{array}
\eqno{(18)}
$$
where  $n=\mid\Phi(\vec{r})\mid^2$ and
$$
L=-\frac{ \hbar^2}{2 m} \nabla^2
+V_{ext}-\mu +\frac{\partial( n \epsilon(n))}{\partial n}+\frac{\partial^2(n\epsilon(n))}{\partial n^2}n.
\eqno{(19)}
$$
Setting $f_\pm$ to zero in Eq.(18), we obtain  coupled equations
$$
\everymath={\displaystyle}
\begin{array}{rcl}
Lu+\frac{\partial^2(n\epsilon(n))}{\partial n^2} \Phi^2 v=\hbar \omega u,\\ \\
Lv+\frac{\partial^2(n\epsilon(n))}{\partial n^2} \Phi^{\ast 2}u=-\hbar \omega v,
\end{array}
\eqno{(20)}
$$
which can be used to calculate the energies $\mathcal{E}=\hbar \omega$ of the elementary excitations.

Now we describe the application of the time-dependent equation (6) to the
case of  
nonlinear dynamics. We turn to the limit of very strong coupling between 
the  interacting bosons in $1D$, the so-called  Tonks-Girardeau gas [27]. In this
 impenetrable bosons case, the energy density $\epsilon(n)$, Eq.(13), reduces
 to $\epsilon(n)=\hbar^2 \pi^2 n^2/6m,$ and Eq.(6) reads [12]
$$
i\hbar \frac{\partial \Psi}{\partial t}=(-\frac{ \hbar^2}{2 m} \frac{\partial^2}{\partial x^2} 
+V_{ext} +\frac{\hbar^2 \pi^2}{2m} \mid\Psi\mid^4)\Psi,
\eqno{(21)}
$$
with
$\int_{-\infty}^{+\infty}\mid \Psi(x,t)\mid^2 dx=N.$

 For  a general time-dependent harmonic trap $V_{ext}=m \omega^2(t) x^2/2,$ with the initial condition $\Psi(x,0)=\Phi(x)$,
where $\Phi(x)$ is the ground-state solution of the time-independent equation
$$
(-\frac{ \hbar^2}{2 m} \frac{\partial^2
}{\partial x^2} +\frac{m \omega^2(0)
x^2}{2}+\frac{\hbar^2 \pi^2}{2m} \mid\Phi\mid^4)\Phi
=\mu \Phi,
\eqno{(22)}
$$
 we show that Eq.(21) reduces to the ordinary differential equations which can provide the exact solution of
 Eq.(21).

Indeed, if we assume that the solution, $\Psi(x,t)$, can be expressed as
$$
\Psi(x,t)=\frac{\Phi(x/\lambda(t))}{\sqrt{\lambda(t)}} e^{-i \beta(t)+i m \frac{x^2}{2 \hbar} \frac{\dot{\lambda}}{\lambda}},
\eqno{(23)}
$$
 we obtain  the following equations for $\lambda$ and $\beta$ after 
inserting Eq.(23) into Eq.(21):
$$
\ddot{\lambda}+ \omega^2(t) \lambda=\frac{\omega^2(0)}{\lambda^3},\:
\lambda(0)=1,\: \dot{\lambda}(0)=0,
\eqno{(24)}
$$
$$
\dot{\beta}=\frac{\mu}{\hbar \lambda^2},\: \beta(0)=0.
\eqno{(25)}
$$
Thus, the ordinary differential equations Eqs.(22), (24), and (25) give the
 exact
 solution of  Eq.(21), and the evolution of the density can be written exactly as 
$$
n(x,t)=(1/\lambda(t))n(x/\lambda(t),0).
\eqno{(26)}
$$

For the case of free expansion, the confining potential is switched off at $t=0$ 
and the atoms fly away. In this case, Eqs.(24) and (25) can be integrated analytically leading to the following solutions for $\lambda$ and $\beta$:
$$
\lambda(t)=\sqrt{1+\omega^2(0) t^2}, \: \beta(t)=\frac{\mu}{\hbar \omega(0)}
\arctan(\omega(0) t).
\eqno{(27)}
$$

We note that self-similar solutions [28] of Eq.(21) were briefly discussed in Ref.[29].

In the large $N$ limit, where the kinetic energy term in Eq.(22) is dropped altogether (the so-called Thomas-Fermi limit), the corresponding density is
$$
n_{TF}(x,t)=\frac{1}{\pi \tilde{\lambda}(t)}[(2N-\frac{x^2}{\tilde{\lambda}^2(t)})]^{1/2}\theta(
2N-\frac{x^2}{\tilde{\lambda}^2(t)}),
\eqno{(28)}
$$
and for the Fourier transform $n(k,t)=\frac{1}{\sqrt{2 \pi}} \int_{-\infty}^{+\infty}
n(x,t) e^{i k x} dx$ we have
$$
n_{TF}(k,t)=\frac{N}{\sqrt{2 \pi}} \frac{2 J_1(\sqrt{2 N} \tilde{\lambda}(t) k)}
{\sqrt{2N} \tilde{\lambda}(t) k},
\eqno{(29)}
$$
where $\tilde{\lambda}(t)=[\hbar/(m \omega(0))]^{1/2} \lambda(t)$ and $J_1$ is the Bessel function of the first order.

The exact many-body wave function, $\Psi_B(x_1,x_2,...,x_N,t),$ of a system of $N$ impenetrable bosons in a time dependent $1D$ harmonic trap,  can be found from the Fermi-Bose mapping [15]
$$
\mid\Psi_B(x_1,x_2,...x_N,t)\mid=\mid \Psi_F(x_1,x_2,...x_N,t)\mid,
\eqno{(30)}
$$
where $\Psi_F$ is the fermionic solution of the time-dependent many-body 
Schr\"odinger equation
$$
i\hbar\frac{\partial \Psi_F}{\partial t}=\sum_{i=1}^{N}(-\frac{\hbar^2}{2m}
\frac{\partial^2}{\partial x_i^2}+\frac{m \omega^2(t) x_i^2}{2})\Psi_F.
\eqno{(31)}
$$
with initial condition $\Psi_F(x_1,x_2,...x_N,0)=\Phi_F(x_1,x_2,..x_N)$,
where $\Phi_F(x_1,x_2,..x_N)$ is the fermionic ground-state solution of the
time-independent Schr\"odinger equation
$$
\sum_{i=1}^{N}(-\frac{\hbar^2}{2m}
\frac{\partial^2}{\partial x_i^2}+\frac{m \omega^2(0) x_i^2}{2})\Phi_F=
E\Phi_F.
$$
Therefore, for the exact density $n_B(x,t)=\int_{-\infty}^{+\infty}dx_2...\int_{-\infty}^{+\infty}dx_N \mid \Psi_B(x,x_2,...x_N,t)\mid^2,$ we have
$$
n_B(x,t)=\frac{1}{\tilde{\lambda}(t)}\sum_{i=0}^{N-1}\mid \phi_i(\frac{x}{\tilde{\lambda}(t)})\mid^2,
\eqno{(32)}
$$
where $\phi_i(x)=c_i \exp(-x^2/2) H_i(x)$, $ c_i=\pi^{-1/4}(2^i i!)^{-1/2}$, and $H_i(x)$ are  Hermite polynomials.
Note that the evolution of $n_B(x,t)$ can be written as Eq.(26), corresponding to a time-dependent dilatation of length scale.

From the knowledge of $n_B(x,t)$ and $n_{TF}(x,t)$ one can evaluate the radii $r(t)=$ 

\noindent
$(\int_{-\infty}^{+\infty}n_B(x,t) x^2 dx)^{1/2}$ and $r_{TF}(t)=(\int_{-\infty}^{+\infty}n_{TF}(x,t) x^2 dx)^{1/2}$ and ratio $r(t)/r_{TF}(t)$. This quantity is equal to 1 at any $t$ for any $N$. This circumstance explains why for a harmonic trap 
the ground-state density profile from Eq.(21) agrees well with the many-body results for systems with a rather small number of atoms $N\approx 10$ [12].
As for a general trap potential we expect such agreement for much larger $N$.

Using the following relation [30]
$$
\sum_{m=0}^n(2^m m!)^{-1} [H_m(x)]^2=(2^{n+1} n!)^{-1}\{[H_{n+1}(x)]^2-H_n(x) H_{n+2}(x)\},
\eqno{(33)}
$$
we obtain an analytical formula for exact density $n_B(x,t)$
$$
n_B(x,t)=\frac{1}{2 \tilde{\lambda}(t)} c_{N-1}^2 e^{-x^2/\tilde{\lambda}^2(t)}
\{[H_N(x/\tilde{\lambda}(t))]^2-H_{N-1}(x/\tilde{\lambda}(t)) H_{N+1}(x/\tilde
{\lambda}(t))\}.
\eqno{(34)}
$$
Then the Fourier transform is given by
$$
n_B(k,t)=\frac{1}{\sqrt{2 \pi}} e^{-\tilde{\lambda}^2(t) k^2/4}
[N L^{(0)}_N(\frac{\tilde{\lambda}^2(t) k^2}{2})+\frac{\tilde{\lambda}^2(t) k^2}{2} L^{(2)}_{N-1}(\frac{\tilde{\lambda}^2(t) k^2}{2})],
\eqno{(35)}
$$
 where  $L^{(\alpha)}_n$ are Laguerre polynomials. Using the asymptotic formula of Hilb's type for the Laguerre polynomial [30], we have the asymptotic
behavior of $n_B(k,t)$ as $N \rightarrow \infty$
$$
n_B(k,t)=\frac{N}{\sqrt{2 \pi}} \frac{2 J_1(\sqrt{2 N} \tilde{\lambda}(t) k)}
{\sqrt{2N} \tilde{\lambda}(t) k}+O(N^{1/4}),
\eqno{(36)}
$$
which is valid uniformly in any bounded region of $k \tilde{\lambda}(t)$.
Eq.(36) for the case of $t=0$ is a rigorous justification of the Thomas-Fermi
approximation [13,31] to a system of non-interacting $1D$ spinless fermions in 
harmonic trapping potentials.

Comparison of Eq.(36) with Eq.(29) shows that in the large $N$ limit 
the KS  like
time-dependent theory for $1D$ impenetrable bosons in a 
time-dependent
harmonic trap, Eq.(29), gives the same result  as the exact many-body
 treatment, Eq.(36). Hence, we have rigorously proved that Eq.(29) correctly describes properties of $1D$ Bose gas in a time-dependent harmonic trap in the limit of large $N$.

In conclusion, we have developed a time-dependent KS like theory for bosons in three and lower-dimensional traps and have obtained coupled equations which can be used to calculate the energies of elementary excitations.  For 
one-dimensional condensate of impenetrable bosons in a general time-dependent
 harmonic trap, it is shown that the corresponding equation reduces to the ordinary differential equations and  gives the same results as the exact
 many-body treatment in the large $N$ limit.

We thank B. Tanatar for his interest and comments and E.B. Kolomeisky for informing us about Ref.[29] and for useful suggestions.
\pagebreak

{\bf References}
\vspace{8pt}

\noindent
1. {\it Bose-Einstein Condensation in Atomic Gases}, Proceedings of the International School of Physics ``Enrico Fermi", edited by M Inquscio, S. Stringari, and C.E. Wieman (IOS Press, Amsterdam, 1999); http://amo.phy.gasou.edu/bec.html; http://jilawww.colorado.edu/bec/   
 and references therein.

\noindent
2. A.L. Fetter and A.A. Svidzinsky, J.Phys.: Condens. Matter { \bf13}, R135 
(2001);
 K. Burnett, M. Edwards, and C.W. Clark, Phys. Today, {\bf 52}, 37 (1999);
 F. Dalfovo {\it et al.}, Rev. Mod. Phys.
 {\bf 71}, 463 (1999);
S. Parkins, and D.F. Walls, Phys. Rep. {\bf 303}, 1 (1998).

\noindent
3.  L.P. Pitaevskii, Zh. Eksp. Teor. Fiz. {\bf 40}, 646 (1961) [Sov. Phys. JETP
 {
\bf 13}, 451 (1961)];
   E.P. Gross, Nuovo Cimento {\bf 20}, 454 (1961); J. Math. Phys. {\bf 4}, 195 (
1963).

\noindent
4.  A. G\"orlitz
{\it et al.},
 Phys. Rev. Lett. {\bf 87}, 130402 (2001).

\noindent
5.  F. Schreck
{\it et al.}, Phys. Rev. Lett. {\bf 87}, 08403 (2001).

\noindent
6. M. Key {\it et al.}, Phys. Rev. Lett. {\bf 84}, 1371 (2000);
 V. Vuleti\'{c} {\it et al.}, Phys. Rev. Lett. {\bf 81}, 5768 (1998);
{\bf 82}, 1406 (1999); {\bf 83}, 943 (1999);
 J. Denschlag, D. Cassettari, and J. Schmiedmayer, Phys. Rev. Lett. {\bf 82}, 2014 (1999);
 M. Morinaga {\it et al.}, Phys. Rev. Lett. {\bf
83}, 4037 (1999);
 I. Bouchoule {\it et al.}, Phys. Rev. A{\bf 59}, R8 (1999).
 H. Gauk {\it et al.}, Phys.
 Rev. Lett. {\bf 81}, 5298 (1998).

\noindent
7. A.L. Zubarev and Y.E. Kim, Phys. Rev. A {\bf 65}, 035601 (2002).

\noindent
8. V. Dunjko, V. Lorent, and M. Olshanii, Phys. Rev. Lett. {\bf 86}, 5413
 (2001).

\noindent
9. J.C. Bronski
 {\it et al.}, Phys. Rev. E{\bf 64}, 056615 (2001);
 L.D. Carr, K.W. Mahmud, and W.P. Reinhardt, Phys. Rev. A{\bf 64}, 033603 
(2001);
 L.D. Carr, C.W. Clark, and W.P. Reinhardt, Phys. Rev. A{\bf 62}, 063610 (2000
).

\noindent
10. M.D. Girardeau, E.M. Wright, and J.M. Triscari, Phys. Rev. A{\bf 63}, 03361
(2001).

\noindent
11.  D.S. Petrov, G.V. Shlyapnikov, and J.T.M. Walraven, Phys. Rev. Lett. {\bf 85}
, 3745 (2000);
 D.S. Petrov, M. Holzmann, and G.V. Shlyapnikov, Phys. Rev. Lett. {\bf 84}, 2551 (2000).

\noindent
12. E.B. Kolomeisky {\it et al.}, Phys. Rev. Lett. {\bf 85}, 1146 (2000).

\noindent
13. B. Tanatar, Europhys. Lett., {\bf 51}, 261 (2000).

\noindent
14. B. Tanatar and K. Erkan, Phys. Rev. A{\b62}, 053601 (2000).

\noindent
15. M.D. Girardeau and E.M. Wright, Phys. Rev. Lett. {\bf 84}, 5239 (2000).

\noindent
16. M. Olshanii, Phys. Rev. Lett. {\bf 81}, 938 (1998).

\noindent
17. H. Monen, M. Linn, and N. Elstner, Phys. Rev. A{\bf 58}, R3395 (1998).

\noindent
18. See, for example, V.E. Barlette, M.M. Leite, and S.K. Adhikari, Am. J. Phys. {\bf 69}, 1010 (2001);
S.K. Adhikari, Am. J. Phys. {\bf 54}, 362 (1986).

\noindent
19. E.H. Lieb, R. Seiringer, and J. Yngvason, Commun. Math. Phys. 224, {]bf 17} (2001).

\noindent
20. P. Hohenberg and W. Kohn, Phys. Rev. B{\bf 136}, 864 (1964);
 W. Kohn and L.J. Sham, Phys. Rev. A{\bf 140}, 1133 (1965).

\noindent
21. G. S. Nunes, J. Phys. B: At. Mol. Opt. Phys. {\bf 32}, 4293 (1999) and references therein.

\noindent
22. K. Huang and C.N. Yang, Phys. Rev. {\bf 105}, 767 (1957);T,D, Lee and C.N. 
Yang, Phys. Rev. {\bf 105},1119 (1957); T.D. Lee, K. Huang and C.N. Yang, Phys.
Rev. {\bf 106},1135 (1957); S.T. Beliaev, JETP {\bf 7}, 299 (1958); T.T. Wu, 
Phys
. Rev. {\bf 115}, 1390 (1959); N. Hugenholtz and D. Pines, Phys. Rev. {\bf 116},
 489 (1959); E.H. Lieb and J Yngvason, Phys. Rev. Lett. {\bf 80}, 2504 (1998) 
and references therein.

\noindent
23. E.H. Lieb and J. Yngvason, J. Stat. Phys. {\bf103}, 509 (2001) and 
references
 therein.

\noindent
24. E.H. Lieb and W. Limiger, Phys. Rev. {\bf130}, 1605, (1963).

\noindent
25. P.-G. de Gennes, {\it Superconductivity of Metals and Alloys} (Benjamin,
New York 1966).

\noindent
26. M. Edwards {\it et al.}, J. Res. Natl. Inst. Stand. Technol. {\bf 101}, 553 (1966);
A. Ruprecht {\it et al.}, Phys. Rev. A{\bf 54}, 4178 (1996);
M. Edwards {\it et al.}, Phys. Rev. Lett. {\bf 77}, 1671 (1996).

\noindent
27. L. Tonks, Phys. Rev. {\bf 50}, 955 (1936); M. Girardeau, J. Math. Phys. {\bf 1}, 516 (1960).

\noindent
28. G.I. Barenblatt, {\it Scaling, self-similarity, and intermediate asymptotics} (Cambridge University Press, Cambridge 1996).

\noindent
29. E.B. Kolomeisky et al., Phys. Rev. Lett. {\bf 86}, 4709 (2001).

\noindent
30. {\it Higher Transcendental Functions}, Bateman Manuscript Project,
    Editor A. Erd\'{e}lyi, Volume II (McGraw-Hill Book Company, Inc., New York 1953);
Husimi was the first to show that the $1D$ density of $N$ fermions, moving independently in a linear time-independent harmonic trap, can be written in terms of the highest ocupied shell alone. See K. Husimi, Proc. Phys. Math. Soc. Jpn. {\bf22}, 264 (1940). 

\noindent
31. D.A. Butts and D.S. Rokhsar, Phys. Rev. A{\bf 55}, 4346 (1997) and 
references
 therein.

\end{document}